\documentclass[useAMS,usenatbib]{mn2e}
\usepackage{epsfig,amsmath,natbib}

\newcommand{\EGeV}{E_\mathrm{GeV}}

\newcount\listnorom
\newcommand\listromanDE{\global\advance \listnorom by 1
{\lowercase\expandafter{(\romannumeral\listnorom)}\ }}

\def\be{\begin{eqnarray}}
\def\ee{\end{eqnarray}}
\def\lsim{\;\raise0.3ex\hbox{$<$\kern-0.75em\raise-1.1ex\hbox{$\sim$}}\;}
\def\gsim{\;\raise0.3ex\hbox{$>$\kern-0.75em\raise-1.1ex\hbox{$\sim$}}\;}

\newcommand{\beq}{\begin{eqnarray}}
\newcommand{\eeq}{\end{eqnarray}}
\def\lsim{\;\raise0.3ex\hbox{$<$\kern-0.75em\raise-1.1ex\hbox{$\sim$}}\;}
\def\gsim{\;\raise0.3ex\hbox{$>$\kern-0.75em\raise-1.1ex\hbox{$\sim$}}\;}

\def\ergs{\rm ~erg~s^{-1}}

\def\epm{{e$^{\pm}$}}

\def\ergs{\rm ~erg~s^{-1}}

\def\arcsec{\hbox{$^{\prime\prime}$}}

\def \hcm {\hbox {\ifmmode $ atom cm$^{-2}\else atom cm$^{-2}$\fi}}

\def \arcsec {\hbox{$^{\prime\prime}$} }

\def\approxgt{\mathrel{\hbox{\rlap{\lower.55ex \hbox {$\sim$}}
        \kern-.3em \raise.4ex \hbox{$>$}}}}
\def\approxlt{\mathrel{\hbox{\rlap{\lower.55ex \hbox {$\sim$}}
        \kern-.3em \raise.4ex \hbox{$<$}}}}

\def \RXJ1713 {RXJ1713.72--3946 }

\def\apj{ApJ}
\def\mnras{MNRAS}
\def\nat{Nat}

\def\araa{ARA\&A}                
\def\aap{A\&A}                   
\def\apjl{ApJ}                   

\def\ssr{Space Sci. Rev.}

\title{
Intermittent synchrotron emission in the synchrotron cut-off
regime:\\ the origin of  gamma-ray flares in the Crab?}

\title{Twinkling pulsar wind nebulae in the synchrotron cut-off regime and the gamma-ray flares in the Crab Nebula}

\author[A.M. Bykov, G.G. Pavlov, A.V. Artemyev, Yu.A. Uvarov]{A.M.Bykov$^{1}$\thanks{E-mail:byk@astro.ioffe.ru},
G.G.Pavlov$^{2,3}$, A.V.Artemyev$^{4}$,Yu.A.Uvarov$^{1,3}$\\
$^{1}$Ioffe Institute for Physics and Technology, 194021
St.Petersburg, Russia\\
$^{2}$525 Davey Laboratory, Pennsylvania State University,
University Park,
 PA 16802\\
$^{3}$State Politechnical University, St.\ Petersburg, Russia\\
$^{4}$Space Research Institute, Russian Academy of Sciences, Moscow,
Russia}

\begin{document}


\pagerange{\pageref{firstpage}--\pageref{lastpage}} \pubyear{}

\maketitle

\label{firstpage}

\begin{abstract}
Synchrotron radiation of ultra-relativistic particles accelerated in
a pulsar wind nebula may dominate its spectrum up to $\gamma$-ray
energies.
Because of the short cooling time of the $\gamma$-ray emitting \epm,
the $\gamma$-ray emission zone
is in the immediate vicinity of  the acceleration site. The particle
acceleration likely occurs at the termination shock of the
relativistic striped wind, where multiple forced magnetic field
reconnections provide strong magnetic fluctuations facilitating
Fermi acceleration processes. The acceleration mechanisms imply the
presence of stochastic magnetic fields in the particle acceleration
region, which cause stochastic variability of the synchrotron
emission. This variability is particularly strong in the steep
$\gamma$-ray tail of the spectrum, where modest fluctuations of the
magnetic field lead to strong flares of spectral flux. In
particular, stochastic variations of magnetic field, which may lead
to quasi-cyclic $\gamma$-ray flares, can be produced by the
relativistic cyclotron ion instability at the termination shock.
Our model calculations of the spectral and temporal evolution of
synchrotron emission in the spectral cut-off regime demonstrate that
the intermittent magnetic field concentrations dominate the
$\gamma$-ray emission from highest energy electrons and provide
fast, strong variability even for a quasi-steady distribution of
radiating particles. The simulated light curves and spectra can
explain the very strong $\gamma$-ray flares observed in the Crab
nebula and the lack of strong variations at other wavelengths. The
model predicts high polarization in the flare phase, which can be
tested with future polarimetry observations.
\end{abstract}
\begin{keywords}
shock waves --- turbulence--- ISM: supernova
remnants---gamma-rays---supernovae: individual (Crab nebula)
\end{keywords}

\section{Introduction}
\label{s_intro}
Strong
flares
of a few days duration have been
discovered recently
 by the {\sl AGILE} and {\sl Fermi}  $\gamma$-ray
observatories in the Crab nebula at energies above 100 MeV
 \citep[see e.g.][and references
therein]{tavani_sci,fermi11,agile11}. The most striking features of
the flares are the extreme amplitude of the photon flux changes,
especially at energies above the exponential cut-off energy of the
quiescent spectrum, and the fast hour-timescale variability. While
the exponential cut-off energy $E_c$ of a typical quiescent
$\gamma$-ray spectrum is $\sim$ 100 MeV,
 the cut-off energy
of $>500$ MeV was found in the April 2011 flare spectrum.
This value of $E_c$
exceeds the energy $\tilde{E}_c \sim m_ec^2/\alpha \sim$ 100 MeV
(where $\alpha = e^2/\hbar\,c$ is the fine-structure constant),
considered to be the maximal
cut-off energy in the
synchrotron models of
$\gamma$-ray emission from the Crab and other pulsar wind nebulae
(PWNe) \citep[][]{dejagerea96,ucb11,agile11a}.
 The value
$\tilde{E}_c$
is the synchrotron photon energy emitted by an electron
whose energy is such that the synchrotron cooling time is equal to
the characteristic gyration time $\omega_g^{-1}$  \citep[see
e.g.][]{gfr83,dejagerea96,aa96}.
 The particle gyration time is considered to be the fastest
acceleration time in a plasma system with frozen-in magnetic field
of the r.m.s. amplitude  $\langle B^2\rangle^{1/2}$
 that exceeds the electric field
 magnitude.
The value of $\tilde{E}_c$
corresponds to the electron Lorentz factor $\gamma_m$ that can be
found from the
equation $\dot{\cal{E}}_{syn}(\gamma_m) =
\dot{\cal{E}}_{acc}(\gamma_m)$. Both the synchrotron loss rate
$\dot{\cal{E}}_{syn}(\gamma)$ and the electron acceleration rate
$\dot{\cal{E}}_{acc}(\gamma)$ depend on the moments (often just
$\langle B^2\rangle$) of the stochastic magnetic field.

In this work we consider the case when the formation length, $r_f=
m_e\,c^2/e\,\langle B^2\rangle^{1/2}$,
 of incoherent synchrotron radiation is much smaller than the typical
synchrotron cooling and acceleration lengths, while the typical
wavelength $\lambda$ of the fluctuating magnetic field is larger
than $r_f$.
In this case
$\dot{\cal{E}}_{syn}(\gamma_m)$, $\dot{\cal{E}}_{acc}(\gamma_m)$,
 and $E_c$
are determined by the same r.m.s. value of the fluctuating magnetic
field, and the bulk of the relativistic electron distribution may
vary on scales that are much larger than the gyration radius $r_g =
\gamma r_f$. Since the synchrotron emissivity of a power-law
electron distribution with spectral index $p$ is proportional to
$B^{(p+1)/2}$, the local emissivity sharply grows with $B$ for large
$p$ values. This means that  the synchrotron radiation in the
cut-off regime (which corresponds to large effective $p$ values) is
governed  by  high statistical moments of the stochastic magnetic
field distribution, and it is intermittent. The intermittency effect
implies that rare strong peaks of the magnetic field distribution
dominate the synchrotron emission \citep{bue08,bubhk09}. It is
particularly important in the synchrotron cut-off regime, when the
typical size of the distribution of
 radiating
electrons
(the synchrotron cooling length) can be
comparable with the correlation length of strong magnetic field
fluctuations.  For instance, this is expected to be the case in
supernova shells, where magnetic fluctuations are produced by
instabilities of anisotropic distributions at the maximal energy of
particles accelerated in the source \citep[see, e.g.,][and
references therein]{ber11}.

\begin{figure}
\centering {
 \rotatebox{0}{
{\includegraphics[height=7.5cm, width=9.5cm]{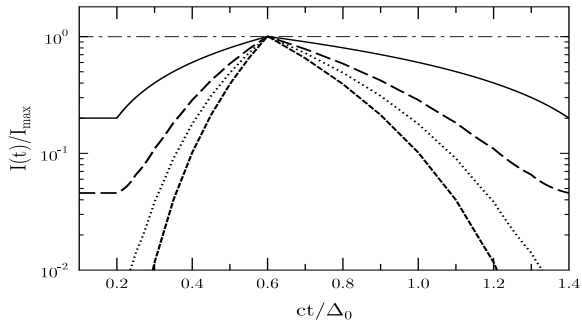} }}} \vskip 5mm
\caption{Light curves of synchrotron emission at 5 keV (dot-dash
line), 500 MeV (dashed line),  1 GeV (dotted line) and 2 GeV
(short-dash line) as a response to an imposed fluctuation with
magnetic field $B(t)$ (solid line) simulated in Model I. The light
curves are normalized to maximal intensity. The background magnetic
field in the emission region was modeled as a stochastic gaussian
field of $\langle B^2(0)\rangle^{1/2} = 0.2$ mG. The imposed
fluctuation $B(t)$ (solid line) is localized in a stripe of a $0.01
\Delta_0$
 thickness
($ \gg r_f)$. The maximum of $B(t)$ = 1 mG is at $ct/\Delta_0$ =
0.6.
 The spatial scale $\Delta_0 \approx 2\times 10^{16}$ cm
for the photon energy $E \approx 1$ GeV provides the photon
variability timescale of about $10^5$ s in the GeV regime.}
\label{light_curve}
\end{figure}

\begin{figure}
\centering {
 \rotatebox{0}{
{\includegraphics[height=7.5cm, width=9.5cm]{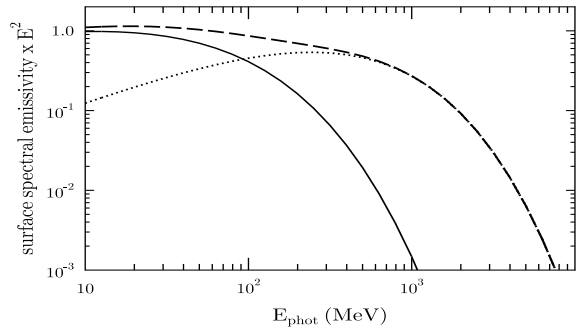} }}}
\caption{Normalized
spectra  of synchrotron radiation at two different time moments
$ct/\Delta_0$ = 0.2 (solid line)
and 0.6 (dashed line), which model the quiescent and flare spectra,
respectively (see Fig.~\ref{light_curve}).
The dotted curve shows the contribution of the variable magnetic
field. The
 power emitted in the GeV flare is about 2$\times 10^{36}
\ergs$, for the Crab parameters.} \label{spectra}
\end{figure}

Since the source emission in the cut-off regime is dominated by just
a single (or a few strongest) concentration(s) of the stochastic magnetic
field, the light curve of the source
in this regime
reflects the lifetime of the magnetic
concentrations rather than the electron acceleration/losses timescales.
 Fast temporal variations will appear even for a quasi-steady
electron distribution. The light curve and the spectral behaviour in
the synchrotron cut-off regime are determined by statistical
characteristics of the magnetic field, which can be described by the
probability distribution function (PDF) of magnetic fluctuations,
$P(B)$.

It is worthwhile to note that intermittent  magnetic fields can be
found in quite different circumstances. For instance, non-Gaussian
distributions of fluctuations, exhibiting gradual tails at large
field amplitudes, have been found in the Earth magnetotail after the
current disruption associated with magnetospheric substorms observed
by {\sl Geotail} and {\sl Cluster} satellites. The energy injection
during the substorms feeds an energy cascade to small-scale
fluctuations with the corresponding increase of intermittency
\citep[see, e.g.,][for a
recent review of
magnetospheric observations]{zimbardoea10}.

In this Letter we show that the
 model of synchrotron emission  in fluctuating magnetic fields,
with account for the intermittency in the spectral cut-off regime,
can explain the nature of the $\gamma$-ray flares observed in the
Crab. We demonstrate the effect of intermittency on the GeV regime
emission using two models. In first model, we simulate  the spectra
of accelerated electrons and positrons in a simple kinetic model of
diffusive Fermi acceleration in the termination shock of striped
wind with account for synchrotron losses. Then we construct the GeV
regime flare light curve and spectra by integrating the synchrotron
emissivity of spatially inhomogeneous particle distribution in the
shock downstream with imposed magnetic field variation. Second
model demonstrates the effect of the
magnetic field
PDF shape on the synchrotron photon
spectra in the cut-off regime. The emission is produced by
Fermi-accelerated pairs in the spectral cut-off regime, in which the
acceleration is balanced by synchrotron losses.

\section{Modeling}
The synchrotron origin of the observed gamma-ray flares assumes the
presence of pairs accelerated to PeV energies. The thickness of the
$\gamma$-radiating region depends on the cooling rate in the
magnetic field:
$\Delta \sim \tau_{\rm syn} c\sim 1.5 \times 10^{15}\cdot B_{\rm
mG}^{-3/2}\cdot \EGeV^{-1/2}$ cm,
where $B_{\rm mG}$ is the r.m.s. magnetic field in mG, and $\EGeV$
is the photon energy in GeV. The size of the $\gamma$-ray emitting
region is very small compared to the size of the nebula.  The
thickness of the layer $\Delta$ is estimated using the standard
electron synchrotron cooling time \citep[e.g.][]{rl79} of the
electron radiating photons at the peak of the power spectrum of
synchrotron emission. The particle acceleration mechanism in PWNe is
not fully understood yet, although some basic features and
constraints have been established \citep[see,
e.g.,][]{kc84,arons07,arons11,keshetea09,klp09,lp10,ss11,bt11,ucb11}.

Recent models of particle acceleration in PWNe consider a
relativistic wind  in the equatorial plane with toroidal stripes of
opposite magnetic field polarity, separated by current sheets.
\citet{ss11} modeled particle acceleration and magnetic field
dissipation at the termination shock of a relativistic striped wind
using 2D and 3D particle-in-cell (PIC) simulations. They found a
complex structure of the flow in the vicinity of the termination
shock. In that model, the shock-driven reconnection in the
downstream transfers the magnetic energy of alternating fields of
the striped wind to accelerated relativistic pairs.  The energy
spectra of electrons accelerated in the reconnection region take the
form of power laws, $N(\gamma) \propto \gamma^{-p}$, with spectral
indices $p \sim 1.5$ that match the radio-optical observations of
the Crab nebula \citep[e.g.][]{Bietenholzea97,hester08,arendtea11}.
The accelerated particles can escape ahead of the shock and generate
magnetic fluctuations in the upstream by the filamentation
and/or Weibel
type instabilities. The turbulence generated by the instabilities
can alleviate Fermi-like diffusive process that accelerates X- and
$\gamma$-ray emitting electrons. This is an important finding as it
simultaneously addresses two problems -- the termination shock
formation in magnetized PWN winds and particle
acceleration\footnote{Note that the standard model of diffusive
Fermi acceleration in a transverse relativistic shock in a
non-striped uniform wind encounters problems when applied to PWN
termination shocks, see, e.g., \citet{nop06,plm09,bt11}.}.
Therefore, the combined action of the reconnection processes and
shock acceleration is expected
in
the intense equatorial pulsar wind. Extensive PIC simulations in
a  wide dynamical range
are needed to demonstrate the feasibility of this
approach  and identify the mechanism of particle acceleration to PeV
energies and their synchrotron emission. However, to construct the
synthetic spectra of the Crab nebula at $\gamma$-ray energies (where
the synchrotron cooling is very important), one needs to account for
the radiative reaction force, which is not yet attainable in the PIC
simulations \citep[][]{ss09,Nishikawaea11}.

Kinetic models can be used to simulate particle acceleration
due to the repetitive interaction of electrons with  magnetic
turbulence in the energetic outflows with account for the synchrotron
cooling.
 The kinetic model
by \citet{bm96} for
 particle
acceleration by both relativistic and transrelativistic shocks,
accompanied by broad dynamic spectra of magnetic fluctuations with
violent motions of relativistic plasma,
predicts a hard broken power-law electron distributions
with slopes $1 \leq p \leq 2$. In this model, particles are
accelerated by strong magnetic fluctuations on timescales comparable
to their gyration period in the r.m.s. magnetic field.

In the simulations described below we complement the kinetic model
for particle acceleration in the vicinity of the striped wind
termination shock with the synchrotron cooling effects to account
for the spectral cut-off regime. To estimate the spectra of
nonthermal leptons accelerated downstream of the PWN termination
shock by Fermi mechanism, one can use a
 Fokker-Planck-type kinetic equation:
\begin{equation}\label{kin_eq}
           \frac{ \partial N }{ \partial t } = k(\gamma) \: \frac{ \partial^2 N}{ \partial z^2 }  -   u\: \frac{ \partial N}{ \partial z } +
      \frac{1}{\gamma^2} \: \frac{\partial}{\partial \gamma} \: \gamma^2 \left[D(\gamma) \:
      \frac{\partial N}{\partial \gamma} + a(\gamma) \:N\right],
\end{equation}
where $z$ is the coordinate along the shock normal\footnote{Since
the scale size $\Delta$ of the PeV electron distribution is much
smaller than the termination shock radius, the problem can be
considered as one-dimensional.}. This  equation, averaged over the
ensemble of strong electromagnetic fluctuations in the vicinity of
the wind termination shock, accounts for diffusion and advection  of
electrons in phase space due to interactions with the fluctuations.
The term with the momentum diffusion coefficient $D(\gamma)$
corresponds to the stochastic Fermi acceleration, $k(\gamma)$ is the
fast particle spatial diffusion coefficient, $u$ is the flow
velocity component along the shock normal,
 and  $a(\gamma)$ is the energy loss rate of an electron
due to synchrotron radiation.

{\sl Model I}. To construct the synchrotron emission spectra and
flaring light curves in the diffusive shock acceleration model, we
simulated spatially inhomogeneous accelerated pair distribution
downstream of the termination shock of the striped wind using
Eq.~\ref{kin_eq}. Short-scale magnetic fluctuations are required to
be present upstream of the shock to allow an efficient diffusive
Fermi acceleration in the transverse relativistic shocks \citep[see
e.g.][and footnote $^1$]{bt11}. Recently, \citet{ss11} have found in
the PIC simulations that the fluctuations can be generated upstream
of the {\em striped wind\/} termination shock. We assume the
fluctuations provide the Bohm diffusion with $k(\gamma) \approx c
r_g(\gamma)/3 $.
 The stochastic Fermi acceleration was neglected in
the model (i.e.,  $D=0$).

\begin{figure}
\centering {
 \rotatebox{0}{
{\includegraphics[height=7.5cm, width=9.5cm]{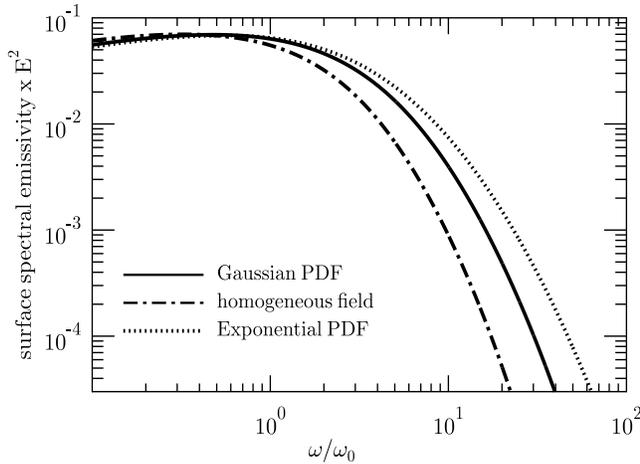} }}}
\caption{
Surface spectral density of synchrotron radiation in the cut-off
regime (in arbitrary units) simulated for different probability
distribution functions of fluctuating magnetic field in the source
(Model II). The characteristic frequency is defined by $\hbar\omega
_0\approx 470 \,\eta$ MeV.} \label{synchr}
\end{figure}

To illustrate the intermittency effect in the cut-off regime, we
simulated a light curve and spectra for a magnetic field fluctuation
imposed in the GeV photon emitting region of scale size $\Delta_0 =
2\times 10^{16}$ cm (for a quiescent magnetic field $B_0$ = 0.2 mG,
downstream of the termination shock). The fluctuation $\delta B(t)$
is localized in a stripe of a $0.01 \Delta_0$ width and has the time
dependence shown by the solid line in Figure~\ref{light_curve}.
Such magnetic field variations may be produced by
 the relativistic ion cyclotron
instability at the termination shock. The instability, proposed by
\citet[][]{sa04} to explain the origin of the optical wisps
in the Crab nebula, can
produce
quasi-cyclic $\gamma$-ray flares in our model.
The scale $\Delta_0$, used in our simulations,
corresponds to about 0.1 of the magnetic field limit cycle found
by \citet[][]{sa04} (see their Figure 2).
 The light curves of the $\gamma$-ray (0.5 , 1 GeV and
2 GeV) and X-ray (5 keV) emission
show the strong response of the $\gamma$-ray emission in the cut-off
regime, while the response is very modest at X-ray energies.
Note that the variability time in the GeV regime
is shorter than the imposed field fluctuation and the cooling time
of PeV electron distribution. This is because the emission in the
cut-off regime is governed by high-order momenta of the magnetic
field. The energy loss rate $a(\gamma)$ that determines the electron
cooling depends on $\langle B^2\rangle$. Therefore, the stochastic
magnetic field realizations in the
emission region
with the same $\langle B^2\rangle$ but different
 high-order momenta (determined by their PDFs) would
correspond to the same particle distributions. However, their photon
spectra
in the cut-off regime are very different. This effect is clearly
seen in Figure~\ref{synchr} discussed below. The synchrotron flares
in the cut-off regime will appear even in the case of steady
electron distributions.

In Figure~\ref{spectra} we show the simulated spectra of
synchrotron emission,
generated downstream of the termination shock,
for $\eta \approx 0.5$, corresponding to quiescent regime (solid
line) and fluctuation maximum (dashed line). They are similar to the
quiescent and flare spectra observed in the Crab nebula by {\sl
AGILE} and {\sl Fermi}  \citep[see][and references
therein]{tavani_sci,fermi11,agile11}.

{\sl Model II.} Apart from the diffusive shock acceleration a
variety of other particle acceleration mechanisms can be important
in the region. The coalescence of magnetic islands, particle
reflection by magnetic islands (both first and second order
Fermi-type processes),
 are expected to be in action, as
it occurs in the Earth magnetosphere
\citep[e.g.][]{drakeea06,zelenyiea10,ss11,ucb11,daughtonea11}. When
the source of strong turbulence is quasi-steady on timescales longer
than $\omega_g^{-1}(\gamma)$, a simple analytical
treatment of the problem
 is also possible. For the case of fast stochastic Fermi acceleration (comparable to the particle
gyration period) by an energetic plasma outflow with strong magnetic
turbulence, the momentum diffusion coefficient in Eq.\ref{kin_eq}
takes the form
 $D(\gamma)
= \gamma^2\, \eta\, \omega_g(\gamma)$,
where $\eta \lsim 1$ is the particle acceleration time measured in
the gyration times.
Note that both $\omega_g(\gamma) = e\,\langle
B^2\rangle^{1/2}/(m_ec\gamma)$ and $a(\gamma)\, = 4r_0^2\langle
B^2\rangle \gamma^2/(9m_ec)$ (where $r_0 = e^2/m_ec^2$) depend on
the same ensemble-averaged value $\langle B^2\rangle $.

The asymptotical shape of the particle spectrum in the cut-off
regime is
\begin{equation}\label{df}
N(\gamma) \propto \gamma^{-p}\,\exp\left[-\int d\gamma\,
a(\gamma)/D(\gamma)\right].
\end{equation}
It is important that, while the
index $p$ depends on
the turbulence spectrum and system geometry, the exponential cut-off
in the particle spectrum is rather universal:
$N(\gamma) \propto \gamma^{-p}\,\exp[-(\gamma/\gamma_0)^2]$, where
$\gamma_0^2 = 9e \eta /(2r_0^2\,\langle B^2\rangle^{1/2})$, i.e.,
$\gamma_0 \approx 5\times 10^9\, \eta\, \langle B_{\rm
mG}^2\rangle^{-1/4}$. The synchrotron emissivity
$\epsilon(\omega,B)$ in a
 local  magnetic field $B$, which is assumed to be uniform on
spatial scales larger than  $r_f
= m_ec^2/eB$, is
given by the equation
\begin{equation}
\epsilon (\omega,B,z)=\frac {\sqrt{3}Be^3}{2\pi mc^2} \int d\gamma\,
\gamma^2 N(z,\gamma)R(\omega /\omega _c)\,,
\end{equation}
where $\omega _c =3eB\gamma^2/2m_ec$ is the characteristic frequency of synchrotron
radiation.
Approximate analytic expressions for the function $R(x)$ were
derived by \citet{cs86} and \cite{za07}.
The spectrum of
synchrotron emission from the downstream region filled with
strong magnetic field fluctuations can be expressed as
\begin{equation}\label{sp}
J(\omega) = \int dB\,dz\, \epsilon (\omega,B,z)P(B).
\end{equation}
%
To illustrate the effect of the magnetic field fluctuations in the
cut-off regime, where high-order statistical moments dominate the
integral in Equation (\ref{sp}), we used the PDF of magnetosonic
type fluctuations, which corresponds to the wisp structures seen in
the polarized optical images presented by \citet{hester08}.
The PDF has the form $P(B) = C_n\, \exp(-b^n/\Theta_n)$, with $n =
1$ and 2, and $b = |B - B_0|/B_0$.
 Here
$C_n$ is the normalization constant, and $\Theta_n$ is the dimensionless width
of the distribution.  The simulated synchrotron
spectra are presented in Figure~\ref{synchr} for the gaussian
($n=2$), exponential ($n=1$), and non-fluctuating magnetic field
distributions.
The characteristic frequency $\omega _0$ for
the
Lorentz factor $\gamma_0$ is given by
 $\hbar\omega _0=27\eta\,m_ec^2/(4\alpha) \approx 470\,\eta$ MeV.
 The
 synchrotron curves in Figure~\ref{synchr} are simulated for $\Theta_n =
 1$ (i.e., for the case of strong fluctuations).
 The results illustrate a strong effect of the PDF shape
on the spectral behaviour in the cut-off
 regime (even for a fixed rms field $\langle B^2\rangle$),
in contrast to a very modest effect in the power-law regime at
$\omega < \omega _0$. Thus, Model II demonstrates that a
reconstruction of the PDF tail of magnetic fluctuations in the
synchrotron emission region (with the same $\langle B^2\rangle$)
would result in a strong change of the synchrotron photon spectrum,
similar to that observed in the GeV flares in the Crab nebula.

\section{Discussion}
In Model I, $\gamma$-ray photons are radiated downstream of the PWN
termination shock by ultrarelativistic electrons.
 The particles are accelerated by some kind of Fermi mechanism in
the vicinity of the termination shock in the striped wind with
reconnecting magnetic fields.
Electrons in the
sharply decreasing
high-energy
tail of the distribution function radiate synchrotron $\gamma$-ray
photons in fluctuating magnetic fields, which results in  flaring
behaviour in the cut-off regime (i.e., at the
high-energy end of the synchrotron
spectrum).

The compressed magnetic field structures, moving through the very
narrow $\gamma$-ray emitting layer just downstream of the
termination shock, result in $\gamma$-ray flares with the strong
spectral variations shown in Figure~\ref{spectra}. The timescale of
the $\gamma$-ray photon variability in the flare is a fraction of
the time $\Delta_0/c \sim 5 \times 10^5$ s
 in the cut-off
regime because of the effect of high-order statistical moments. At
the same time, the variability at energies below the cut-off
energies is slower, and of much smaller amplitudes, as it is clearly
seen from comparison of the light curves at 5 keV and above 500 MeV in
Figure~\ref{light_curve}. This is consistent with the lack of strong
variations in the Crab at the X-ray and optical
energies\footnote{Note that we are discussing variations in the flux
integrated over the downstream region (over $z$ in our 1D model).
The lack of strong variations in the flux does not contradict  local
variations in the synchrotron intensity, such as the moving wisps.}
\citep[][]{tennant11,caraveo10}.
The frequency of the $\gamma$-ray
flares is determined in our model by the frequency of occurrence of
 extreme values of magnetic fluctuations. The strong magnetic
field compressions could be associated with the variable wisps
observed with the quasi-periodic cycle of a year timescale
\citep[e.g.][]{scargle69,hesterea02,Bietenholzea04}.  To model the
dynamics of the wisps, \citet{sa04} simulated outward-propagating
magnetosonic waves downstream of the shock generated by relativistic
cyclotron instability of gyrating ions. The process achieves a limit
cycle, in which the waves are launched with periodicity on the order
of the ion Larmor time (a year timescale). Compressions in the
magnetic field and pair density, associated with these waves, can
reproduce the behaviour of the wisp and ring features. The
high-resolution relativistic MHD models by \citet{camusea09} have
revealed a highly variable structure of the pulsar wind termination
shock with
 quasi-periodic behaviour within the periods of about 2 years, and
MHD turbulence on scales shorter than 1 year.

 The synchrotron
emission in the $\gamma$-ray regime is dominated by infrequent
quasi-coherent structures, and therefore it should be highly
polarized, that can be tested with future time resolved polarimetry
observations.

The strong $\gamma$-ray variability predicted by our model is
consistent with that observed in the Crab nebula. No simultaneous
strong flares at other wavelengths is expected in our model. The
$\gamma$-ray flares can occur even  for a steady electron
distribution function with the maximal energies not exceeding the
limit of particle acceleration by the Fermi mechanism with strong
magnetic fluctuations at the extended wind termination shock.  This
is the main difference from the model by \citet[][]{yuanea11}, which
explains the flares by varying maximal electron energies  in
isolated knots that must transfer a substantial power (above
10$^{36} \ergs$) to the observed GeV photons. The variability of the
maximal energies of the electrons would result in simultaneous GeV
and 100 TeV regime flares since the electrons producing the
synchrotron GeV photons radiate also 100 TeV regime photons by
inverse Compton scattering. Also, \citet{bi11} pointed out that the
TeV photons produced by inverse Compton scattering of soft radiation
by the variable distribution of accelerated electrons should be
variable on timescales similar to those observed at GeV energies by
{\sl AGILE} and {\sl Fermi-LAT}.

Recently, \citet{kl11} proposed a model in which the Crab flares
originate in the inner knot (within about $1\arcsec$ from the Crab
pulsar) with strong Doppler beaming effects. In our model the region
of GeV photon emission  is about 10$\arcsec$ away from the pulsar.
The Doppler-boosted synchrotron emission  from a corrugated shock,
proposed by \citet{lbm11}, would be accompanied by flares in the
cut-off regime of the inverse Compton TeV photons. This is different
from the prediction of our model, in which the strong variability of
GeV synchrotron emission is due to magnetic field variability,
strongly amplified in the cut-off regime, while the amplitude of TeV
emission variation is expected to be much less prominent.

We thank the referee J.Arons for constructive comments.
 A.M.B, G.G.P and Y.A.U were supported in part by
 the Russian government grant
11.G34.31.0001 to the Saint-Petersburg State Politechnical
University, and also by the RAS Programs and by the RFBR grant
11-02-12082-ofi-m-2011. The numerical simulations were performed at
JSCC RAS and the SC at Ioffe Institute. The research by A.M.B was
supported in part by the National Science Foundation under Grant
PHY05-51164 and by ISSI (Bern). G.G.P was supported in part by NASA
grants NNX09AC84G and NNX09AC81G, and NSF grant AST09-08611.

\bibliographystyle{mn2e}



\end{document}